\documentclass{JHEP3}

\usepackage{epsfig,multicol,axodraw}

\newcommand{\I}{{\rm i}}
\newcommand{\bk}{{\bf k}}
\newcommand{\bp}{{\bf p}}
\newcommand{\x}{{\bf x}}

\newcommand{\sG}{{\sf G}}

\newcommand{\T}{{\sf T}}

\newcommand{\D}{{\rm d}}
\newcommand{\ubar}{{\bar u}}

\newcommand{\wl}{\omega_L}

\newcommand{\cw}{c_{\mathrm{w}}}
\newcommand{\sw}{s_{\mathrm{w}}}

\newcommand{\diagTree}{
\begin{picture}(140,60)(-70,-30)
\ArrowLine(-60,-25)(0,-25)
\Text(-56,-29)[t]{$f\,(2)$}
\Vertex(0,-25){2}
\ArrowLine(0,-25)(60,-25)
\Text(56,-29)[t]{$f\,(4)$}
\Photon(0,-25)(0,25){3}{3.5}
\Text(6,0)[l]{$Z$}
\Vertex(0,25){2}
\ArrowLine(-60,25)(0,25)
\Text(-56,29)[b]{$\nu_l\,(1)$}
\Vertex(0,25){2}
\ArrowLine(0,25)(60,25)
\Text(56,29)[b]{$\nu_l\,(3)$}
\end{picture}}

\newcommand{\diagTreeNNt}{
\begin{picture}(140,60)(-70,-30)
\ArrowLine(-60,-25)(0,-25)
\Text(-56,-29)[t]{$\nu_{l^\prime}$}
\Vertex(0,-25){2}
\ArrowLine(0,-25)(60,-25)
\Text(56,-29)[t]{$\nu_{l^\prime}$}
\Photon(0,-25)(0,25){3}{3.5}
\Text(6,0)[l]{$Z$}
\Vertex(0,25){2}
\ArrowLine(-60,25)(0,25)
\Text(-56,29)[b]{$\nu_l$}
\Vertex(0,25){2}
\ArrowLine(0,25)(60,25)
\Text(56,29)[b]{$\nu_l$}
\end{picture}}

\newcommand{\diagTreeNNu}{
\begin{picture}(140,60)(-70,-30)
\ArrowLine(-60,-25)(0,-25)
\Text(-56,-29)[t]{$\nu_l$}
\Vertex(0,-25){2}
\Line(0,25)(18,10)
\ArrowLine(18,10)(60,-25)
\Text(56,-29)[t]{$\nu_l$}
\Photon(0,-25)(0,25){3}{3.5}
\Text(6,0)[l]{$Z$}
\Vertex(0,25){2}
\ArrowLine(-60,25)(0,25)
\Text(-56,29)[b]{$\nu_l$}
\Vertex(0,25){2}
\Line(0,-25)(18,-10)
\ArrowLine(18,-10)(60,25)
\Text(56,29)[b]{$\nu_l$}
\end{picture}}

\newcommand{\diagIa}{
\begin{picture}(140,80)(-70,-35)
\ArrowLine(-60,-30)(0,-30)
\Text(-56,-34)[t]{$f$}
\Vertex(0,-30){2}
\ArrowLine(0,-30)(60,-30)
\Text(56,-34)[t]{$f$}
\Photon(0,-30)(0,5){3}{2.5}
\Text(6,-20)[l]{$Z$}
\Vertex(0,5){2}
\ArrowLine(-60,5)(-22,5)
\Text(-56,9)[b]{$\nu_l$}
\Vertex(-22,5){2}
\ArrowLine(-22,5)(0,5)
\Text(-9,1)[tr]{$l$}
\ArrowLine(0,5)(22,5)
\Text(9,1)[tr]{$l$}
\Vertex(22,5){2}
\ArrowLine(22,5)(60,5)
\Text(56,9)[b]{$\nu_l$}
\PhotonArc(0,5)(22,0,180){3}{6}
\Text(0,35)[b]{$W$}
\end{picture}}

\newcommand{\diagIb}{
\begin{picture}(140,80)(-70,-35)
\ArrowLine(-60,-30)(0,-30)
\Text(-56,-34)[t]{$f$}
\Vertex(0,-30){2}
\ArrowLine(0,-30)(60,-30)
\Text(56,-34)[t]{$f$}
\Photon(0,-30)(0,5){3}{2.5}
\Text(6,-20)[l]{$Z$}
\Vertex(0,5){2}
\ArrowLine(-60,5)(-22,5)
\Text(-56,9)[b]{$\nu_l$}
\Vertex(-22,5){2}
\ArrowLine(-22,5)(0,5)
\Text(-9,1)[tr]{$l$}
\ArrowLine(0,5)(22,5)
\Text(9,1)[tr]{$l$}
\Vertex(22,5){2}
\ArrowLine(22,5)(60,5)
\Text(56,9)[b]{$\nu_l$}
\DashCArc(0,5)(22,0,180){4}
\Text(0,35)[b]{$\phi$}
\end{picture}}

\newcommand{\diagIIa}{
\begin{picture}(140,80)(-70,-40)
\ArrowLine(-60,-30)(0,-30)
\Text(-56,-34)[t]{$f$}
\Vertex(0,-30){2}
\ArrowLine(0,-30)(60,-30)
\Text(56,-34)[t]{$f$}
\Photon(0,-30)(0,0){3}{3}
\Text(6,-15)[l]{$Z$}
\Vertex(0,0){2}
\ArrowLine(-60,30)(-25,30)
\Text(-56,34)[b]{$\nu_l$}
\Vertex(-25,30){2}
\ArrowLine(-25,30)(25,30)
\Text(0,35)[b]{$l$}
\Vertex(25,30){2}
\ArrowLine(25,30)(60,30)
\Text(56,34)[b]{$\nu_l$}
\Photon(0,0)(-25,30){3}{3}
\Text(-17,11)[r]{$W$}
\Photon(0,0)(25,30){3}{3}
\Text(19,11)[l]{$W$}
\end{picture}}

\newcommand{\diagIIb}{
\begin{picture}(140,80)(-70,-40)
\ArrowLine(-60,-30)(0,-30)
\Text(-56,-34)[t]{$f$}
\Vertex(0,-30){2}
\ArrowLine(0,-30)(60,-30)
\Text(56,-34)[t]{$f$}
\Photon(0,-30)(0,0){3}{3}
\Text(6,-15)[l]{$Z$}
\Vertex(0,0){2}
\ArrowLine(-60,30)(-25,30)
\Text(-56,34)[b]{$\nu_l$}
\Vertex(-25,30){2}
\ArrowLine(-25,30)(25,30)
\Text(0,35)[b]{$l$}
\Vertex(25,30){2}
\ArrowLine(25,30)(60,30)
\Text(56,34)[b]{$\nu_l$}
\DashLine(0,0)(-25,30){3}
\Text(-17,11)[r]{$\phi$}
\DashLine(0,0)(25,30){3}
\Text(19,11)[l]{$\phi$}
\end{picture}}

\newcommand{\diagIIc}{
\begin{picture}(140,80)(-70,-40)
\ArrowLine(-60,-30)(0,-30)
\Text(-56,-34)[t]{$f$}
\Vertex(0,-30){2}
\ArrowLine(0,-30)(60,-30)
\Text(56,-34)[t]{$f$}
\Photon(0,-30)(0,0){3}{3}
\Text(6,-15)[l]{$Z$}
\Vertex(0,0){2}
\ArrowLine(-60,30)(-25,30)
\Text(-56,34)[b]{$\nu_l$}
\Vertex(-25,30){2}
\ArrowLine(-25,30)(25,30)
\Text(0,35)[b]{$l$}
\Vertex(25,30){2}
\ArrowLine(25,30)(60,30)
\Text(56,34)[b]{$\nu_l$}
\DashLine(0,0)(-25,30){3}
\Text(-17,11)[r]{$\phi$}
\Photon(0,0)(25,30){3}{3}
\Text(19,11)[l]{$W$}
\end{picture}}

\newcommand{\diagIIIa}{
\begin{picture}(140,80)(-70,-35)
\ArrowLine(-60,-30)(25,-30)
\Text(-56,-34)[t]{$f$}
\Vertex(25,-30){2}
\ArrowLine(25,-30)(60,-30)
\Text(56,-34)[t]{$f$}
\Photon(25,-30)(25,5){3}{2.5}
\Text(31,-12)[l]{$Z$}
\Vertex(25,5){2}
\ArrowLine(-60,5)(-37,5)
\Text(-56,9)[b]{$\nu_l$}
\Vertex(-37,5){2}
\ArrowLine(-37,5)(7,5)
\Text(-15,0)[t]{$l$}
\Vertex(7,5){2}
\ArrowLine(7,5)(25,5)
\ArrowLine(25,5)(60,5)
\Text(56,9)[b]{$\nu_l$}
\PhotonArc(-15,5)(22,0,180){3}{6}
\Text(-15,35)[b]{$W$}
\end{picture}}

\newcommand{\diagIIIb}{
\begin{picture}(140,80)(-70,-35)
\ArrowLine(-60,-30)(25,-30)
\Text(-56,-34)[t]{$f$}
\Vertex(25,-30){2}
\ArrowLine(25,-30)(60,-30)
\Text(56,-34)[t]{$f$}
\Photon(25,-30)(25,5){3}{2.5}
\Text(31,-12)[l]{$Z$}
\Vertex(25,5){2}
\ArrowLine(-60,5)(-37,5)
\Text(-56,9)[b]{$\nu_l$}
\Vertex(-37,5){2}
\ArrowLine(-37,5)(7,5)
\Text(-15,0)[t]{$l$}
\Vertex(7,5){2}
\ArrowLine(7,5)(25,5)
\ArrowLine(25,5)(60,5)
\Text(56,9)[b]{$\nu_l$}
\DashCArc(-15,5)(22,0,180){4}
\Text(-15,35)[b]{$\phi$}
\end{picture}}

\newcommand{\diagIVa}{
\begin{picture}(140,80)(-70,-40)
\ArrowLine(-60,-30)(-25,-30)
\Text(-56,-34)[t]{$f$}
\Vertex(-25,-30){2}
\ArrowLine(-25,-30)(25,-30)
\Text(0,-34)[t]{$f'$}
\Vertex(25,-30){2}
\ArrowLine(25,-30)(60,-30)
\Text(56,-34)[t]{$f$}
\ArrowLine(-60,30)(-25,30)
\Text(-56,34)[b]{$\nu_l$}
\Vertex(-25,30){2}
\ArrowLine(-25,30)(25,30)
\Text(0,35)[b]{$l$}
\Vertex(25,30){2}
\ArrowLine(25,30)(60,30)
\Text(56,34)[b]{$\nu_l$}
\Photon(-25,-30)(-25,30){3}{4}
\Text(-32,0)[r]{$W$}
\Photon(25,-30)(25,30){3}{4}
\Text(32,0)[l]{$W$}
\end{picture}}

\newcommand{\diagIVb}{
\begin{picture}(140,80)(-70,-40)
\ArrowLine(-60,-30)(-25,-30)
\Text(-56,-34)[t]{$f$}
\Vertex(-25,-30){2}
\ArrowLine(-25,-30)(25,-30)
\Text(0,-34)[t]{$f'$}
\Vertex(25,-30){2}
\ArrowLine(25,-30)(60,-30)
\Text(56,-34)[t]{$f$}
\ArrowLine(-60,30)(-25,30)
\Text(-56,34)[b]{$\nu_l$}
\Vertex(-25,30){2}
\ArrowLine(-25,30)(25,30)
\Text(0,35)[b]{$l$}
\Vertex(25,30){2}
\ArrowLine(25,30)(60,30)
\Text(56,34)[b]{$\nu_l$}
\Photon(-25,-30)(25,30){3}{5}
\Text(-22,0)[r]{$W$}
\Photon(25,-30)(-25,30){3}{5}
\Text(22,0)[l]{$W$}
\end{picture}}

\title{Flavour-dependent radiative correction to
neutrino--neutrino refraction}

\author{Alessandro Mirizzi$^1$, Stefano Pozzorini$^2$,
    Georg G.~Raffelt$^1$, Pasquale D.~Serpico$^2$\\
$^1$Max-Planck-Institut f\"ur Physik
(Werner-Heisenberg-Institut), F\"ohringer Ring 6, 80805~Munich,
Germany\\
$^2$Physics Department, Theory Group,
CERN, CH-1211 Geneva 23, Switzerland\\
E-mails: \email{amirizzi@mppmu.mpg.de},
\email{stefano.pozzorini@cern.ch},
\email{raffelt@mppmu.mpg.de},
\email{serpico@cern.ch}}

\preprint{CERN-PH-TH/2009-127\\ MPP-2009-81}

\abstract{In the framework of the Standard Model we calculate the
  flavour non-universal correction for neutrino refraction in a
  neutrino background and verify a similar previous result for the
  case of ordinary-matter background. The dominant term arises at loop
  level and involves $\tau$ leptons circulating in the loop. These
  $\mathcal{O}(G_{\rm F}m_\tau^2)$ corrections to the tree-level
  potential provide the dominant refractive difference between
  $\nu_\mu$ and $\nu_\tau$ unless the medium contains $\mu$ or
  $\tau$ leptons. Our results affect the flavour evolution of dense
  neutrino gases and may be of interest for collective three-flavour
  oscillations of supernova neutrinos.  We spell out explicitly how
  these non-universal neutrino--neutrino interactions enter the flavour
  oscillation equations.}

\keywords{Neutrino Physics, Standard Model, Supernovae}

\begin{document}

\section{Introduction}       \label{sec:introduction}

The dispersion relation of any particle is modified by a background
field or medium. For neutrinos this effect is extremely faint, yet the
matter effect is crucial for neutrino oscillations where the feeble
refractive indices compete with the minuscule mass
differences~\cite{Wolfenstein:1977ue, Mikheev:1986gs}.

Since neutral-current interactions among the three active neutrinos
are flavour universal, they can be usually ignored for neutrino
oscillation phenomenology, where only the relative energy shift
between flavours is important. Moreover, energy scales of
interest are well below the weak boson mass $M_W$, so for low-energy
neutrinos in ordinary matter (no $\mu$ or $\tau$ leptons) the
potential-energy shift between flavours can be calculated simply from
the effective charged-current Hamiltonian density
\begin{equation}
{\mathcal H}_{\mathrm{MSW}}=
\frac{G_{\rm F}}{\sqrt{2}}[\bar{e}\gamma_\alpha(1-\gamma_5) \nu_e]
[\bar{\nu}_e\gamma^\alpha(1-\gamma_5) e]=
\frac{G_{\rm F}}{\sqrt{2}}[\bar{\nu}_e\gamma_\alpha(1-\gamma_5) \nu_e]
[\bar{e}\gamma^\alpha(1-\gamma_5) e]\,,\label{hamcc}
\end{equation}
where the second equality follows from the usual Fierz
rearrangement. One thus finds the well-known tree-level energy
splitting for electron neutrinos with respect to $\nu_\mu$ or
$\nu_\tau$,
\begin{equation}\label{eq:basicshift}
\Delta E_{\nu_e\nu_{\mu,\tau}}=E_{\nu_e}-E_{\nu_{\mu,\tau}}
=\sqrt{2}\,G_{\rm F}n_e\,,
\end{equation}
where $n_e$ is understood as net density of electrons minus
positrons. In lowest order of perturbation theory, the $\nu_\mu$ and
$\nu_\tau$ dispersion relations coincide and $\Delta
E_{\nu_\mu\nu_\tau}=0$.

Equation~(\ref{eq:basicshift}) provides a sufficient description of
the matter effect in most practical cases, most notably solar
neutrinos, but care must be taken in some circumstances where either
the leading term vanishes ($n_e\simeq 0$) or subtle three-flavour
effects come into play. For example, it has long been
appreciated that radiative corrections break the
$\nu_\mu$--$\nu_\tau$ degeneracy in an ordinary medium because the
charged-lepton mass ($m_\mu$ or $m_\tau$) in the loops distinguishes
between flavours~\cite{Botella:1986wy}.
In the low-energy limit, the loop-induced
$\nu_\mu$--$\nu_\tau$ energy shift has the same structure as
Eq.~(\ref{eq:basicshift}) and thus can be represented by an effective
net tau-lepton density in the form
\begin{equation}\label{eq:botella0a}
\Delta E_{\nu_\tau\nu_\mu}=E_{\nu_\tau}-E_{\nu_\mu}
=\sqrt{2}\,G_{\rm F}\,n_\tau^{\rm eff}\,.
\end{equation}
Botella, Lim and
Marciano~\cite{Botella:1986wy} studied explicitly the case of an
ordinary
electrically neutral medium and found the equivalent of
\begin{equation}\label{eq:botella0}
n_\tau^{\rm eff} = \epsilon\,
\frac{3}{2}
\left[\log\left(\frac{M_W^2}{m_\tau^2}\right)-1+\frac{Y_n}{3}\right]
n_B\,,
\end{equation}
where $n_B=n_p+n_n$ is the baryon density and $Y_n=n_n/n_B$ the
neutron/baryon fraction, the remainder being protons. We have
introduced the small parameter
\begin{equation}
\epsilon = \frac{G_{\rm F}m^2_\tau}{\sqrt2\,\pi^2}
= 2.64\times 10^{-6}
\label{epsi}
\end{equation}
that will appear in all of our results.  Here and in the following we
neglect $m_e$, $m_\mu$, the first-generation quark masses and the
neutrino energy relative to $m_\tau$ and $M_W$.

The quantity $Y_n/3$ in Eq.~(\ref{eq:botella0}) is at most a 5\%
correction. Neglecting it provides
\begin{equation}\label{eq:neffbaryon}
n_\tau^{\rm eff}\simeq 2.6\times10^{-5}\,n_B\,.
\end{equation}
The only known example where this small density can be of practical
interest is the case of supernova (SN) neutrinos propagating in a
dense environment after leaving the neutrino sphere of the collapsed
core. At large distances they encounter two resonances where the usual
matter effect causes a level crossing with the ``atmospheric'' and
``solar'' mass differences, respectively~\cite{Dighe:1999bi}. Much
closer to the core the radiative effect will cause similar
resonances~\cite{Akhmedov:2002zj}. However, the $\nu_\mu$,
$\bar\nu_\mu$, $\nu_\tau$ and $\bar\nu_\tau$ fluxes and spectra are
probably equal, so one may think that $\nu_\mu$--$\nu_\tau$
transformations play no role.

This picture has radically changed with the insight that typically
flavour oscillations do occur close to the neutrino sphere. The
refractive effect of neutrinos on neutrinos is not fully captured by
a refractive energy shift because if neutrinos oscillate, the same
applies to the neutrino background, providing a ``flavour
off-diagonal refractive index''~\cite{Pantaleone:1992eq,
Sigl:1992fn}. In other words, the neutrino oscillation equations
become intrinsically nonlinear, leading to collective forms of
oscillation~\cite{Kostelecky:1994dt, Pastor:2001iu}, a subject of
intense recent investigation (see the recent
review~\cite{Duan:2009cd} and references therein). Since collective
oscillations operate close to the neutrino sphere, here the
$\nu_\mu$--$\nu_\tau$ resonance can become important in a
three-flavour treatment~\cite{EstebanPretel:2007yq,Gava:2008rp}. On the other
hand, collective oscillations are suppressed if the matter effect is
too large, typically when~$n_B\gtrsim
n_\nu$~\cite{EstebanPretel:2008ni}. The SN matter profile and
neutrino flux vary with time and depend strongly on the properties
of the progenitor star, so depending on circumstances $n_\nu$ can be
larger or smaller than $n_B$ in the collective oscillation region a
few hundred kilometers above the neutrino sphere. When $n_\nu
\gtrsim n_B$ and non-linear effects become important, it is
reasonable to expect that the leading effect breaking the degeneracy
between $\nu_\mu$ and $\nu_\tau$ is actually provided by {\it
loop-induced effects due to the neutrino background itself}. In this
article we fill a gap in the literature and calculate this
non-universal radiative correction for neutrino refraction in a
neutrino background.

Our calculation is described in Sec.~\ref{calc} where we begin with
a background of charged fermions, confirming the original result
shown in Eq.~(\ref{eq:botella0}). From this starting point we
include neutrinos as a background and derive our main result, an
effective neutrino--neutrino Hamiltonian that includes $m_\tau$
effects in the loop. In Sec.~\ref{formalism} we spell out how to
include our result in the non-linear flavour oscillation equations
before concluding in Sec.~\ref{concl}.

\section{Calculation of the radiative corrections}   \label{calc}

The energy shift between different neutrino species in a given
medium is proportional to the difference of forward scattering
amplitudes~\cite{Raffelt:1996wa,Fukugita:2003en}. Since we deal with
energies well below the electroweak scale, the results of the
one-loop calculation within the electroweak theory can be expressed
in terms of an effective four-fermion Hamiltonian.  Generically, the
quantum effects manifest themselves in the low-energy theory as
corrections to the coefficients of the operators. (In addition they
could induce higher-order operators mediating processes forbidden in
the original theory.) The corrections, however, need not respect the
original symmetries of the interaction. The case at hand makes no
exception: we shall see by explicit calculation that at one loop,
although the Lorentz structure of the tree-level current-current
interaction is preserved, flavour universality of the ``neutral
currents'' breaks down.

\subsection{Tree level}

To set up our problem and fix the notation, we recall that neutrinos
interact with matter via the charged  and neutral weak-current
Lagrangian densities
\begin{eqnarray}
{\mathcal L}_{\rm{CC}}&=&-\frac{g}{2\sqrt{2}}(J^{\rm{CC}}_\alpha W^{+\alpha}
+{J_\alpha^{\rm{CC}}}^\dagger W^{-\alpha})\,,\label{lagcc1}\\
{\mathcal L}_{\rm{NC}}&=&-\frac{g}{2\cw}\,J^{\rm{NC}}_\alpha Z^{\alpha}\,.\label{lagcc2}
\end{eqnarray}
Here, $g$ denotes the SU(2) weak coupling, $\cw=\sqrt{1-\sw^2}$ is the
cosine of the weak mixing angle, and $W^{\pm}$ and $Z$ represent the
charged and neutral gauge bosons with masses $M_W$ and $M_Z=M_W/\cw$,
respectively.  These parameters are related to the Fermi constant
$G_{\rm F}$ and the fine structure constant $\alpha$ by
\begin{equation}
G_{\rm F}=\frac{\sqrt{2}\,g^2}{8\,M_W^2}=\frac{\pi\alpha}{\sqrt2
\sw^2\,M_W^2}\,.
\end{equation}
In terms of the weak-interaction eigenstates, the  charged
current is
\begin{eqnarray}
{J_\alpha^{\rm{CC}}}&=&
\sum_{l}\bar{\nu}_l\gamma_\alpha(1-\gamma_5)l+\sum_i\bar{u}_i
\gamma_\alpha(1-\gamma_5) d_i
\label{ccur}
\end{eqnarray}
while the neutral current is
\begin{eqnarray}
J^{\rm{NC}}_{\alpha}&=& \sum_{l}\left[
\bar{\nu}_l\gamma_\alpha(c_V^{\nu}-c_A^{\nu}\gamma_5)\nu_l
+\bar{l}\gamma_\alpha(c_V^{l}-c_A^{l}\gamma_5)l\right]+
\nonumber\\
&+&\,\sum_i\left[\bar{u}_i\gamma_\alpha(c_V^{u}-c_A^{u}\gamma_5) u_i
+\,\bar{d}_i\gamma_\alpha(c_V^{d}-c_A^{d}\gamma_5) d_i\right].\label{ncur}
\end{eqnarray}
The indices $l$ and $i$ run over the lepton and quark generations,
respectively. The vector and axial couplings of a fermion
$f=\nu$, $e$, $u$ or $d$ are
\begin{equation}
 c_V^{f}=T_{3f}-2 Q_{f}\sw^2\,,\qquad
 c_A^{f}=T_{3f}\,,
\label{VAcouplings}
\end{equation}
where $Q_{f}=\left\{0,-1,\frac{2}{3},-\frac{1}{3}\right\}$ is the
electric charge and $T_{3f}=\left\{\frac{1}{2},
-\frac{1}{2},\frac{1}{2}, -\frac{1}{2}\right\}$ is the third component
of the weak isospin.  In particular $c_V^\nu=c_A^\nu=1/2$ for
neutrinos. We need to calculate the shift in the potential of a
neutrino $\nu_l=\nu_\mu,\nu_\tau$ interacting with a medium consisting
of ordinary matter, i.e.~fermions $f=u,d,e$.  This interaction is
described by the effective neutral-current Hamiltonian density
\begin{equation}
{\mathcal H}_{\nu_l f}=\frac{G_{\rm F}}
{\sqrt{2}}[\bar{\nu}_l\gamma_\alpha(1-\gamma_5) \nu_l]
[\bar{f}\gamma^\alpha(c_V^f-c_A^f\gamma_5) f] \,.
\label{hamgen}
\end{equation}
Charged-current interactions can be ignored since we consider only the
potential-energy difference between $\mu$-- and $\tau$--neutrinos,
which is a purely radiative effect, and the medium is free from
$\mu$-- and $\tau$--leptons.

Assuming homogeneity, the potential in terms of the Dirac spinors of
the (left-handed) neutrino $u_\nu(p)$ and of the fermion $u_f(k,s)$ is
\begin{equation}
{ V}_{\nu_l f}=\frac{G_{\rm F}}{\sqrt{2}}\sum_s
\int {\rm d}^3 \bk\, {\rm d}^3 \x\,
{ F}_f(\bk,s)\,
[\bar{u}_\nu(p)\gamma_\alpha(1-\gamma_5)u_\nu(p)]
[\bar{u}_f(k,s)\gamma^\alpha(c_V^f-c_A^f\gamma_5) u_f(k,s)]\,
\label{hamgen2}
\end{equation}
where the spatial integral is over the normalization volume
(containing one neutrino), ${ F}_f(\bk,s)$ is the momentum
distribution of the fermion background (normalized to 1), and $s$
denotes the fermion polarization.  The matrix elements are calculated
between identical initial and final states, both in momentum and spin,
i.e.~they correspond to the {\it forward scattering amplitude}.  This
leads to~\cite{Giunti:1990pp}
\begin{eqnarray}
{V}_{\nu_l f}&=&\frac{G_{\rm F}}{\sqrt{2}}\,
\sum_s
\int {\rm d}^3 \bk\,  { F}_f (\bk,s)
\left(\frac{2 p_\nu^\alpha}{E_\nu}\right)
\left[\bar{u}_f(k,s)\gamma_\alpha(c_V^f-c_A^f\gamma_5) u_f(k,s)\right]
\nonumber\\
&=&\sqrt{2}\,G_{\rm F}\,\sum_s\,\int {\rm d}^3 \bk\,  {F}_f(\bk,s)
\left(\frac{p_\nu^\alpha}{E_\nu}\right)
\left( n_f\frac{c_V^f\,k_\alpha-c_A^f\,m_f\,s_\alpha}{E_f}\right)\,,
\label{polpotenatial}
\end{eqnarray}
where $m_f$ is the background fermion mass and $s^\alpha$ its
polarization four vector, such that $s\cdot k=0,\,s^2=-1$.

When the medium consists of unpolarized
ordinary-matter fermions ($f=e,u,d$), the term linear in $s^\alpha$
averages to zero. Thus the neutrino potential is independent of the
axial coupling $c_A^f$ and reads
\begin{equation}
{V}_{\nu_l f}=\sqrt{2}\,G_{\rm F}\,n_f\,c_V^f\,
\sum_s\int {\rm d}^3 \bk\,{F}_f(\bk,s)
\left(1-{\bf v}_\bp\cdot {\bf v}_\bk\right)
=\sqrt{2}\,G_{\rm F}\,n_f\,c_V^f\,,\label{shift}
\end{equation}
where ${\bf v}_\bp$ and ${\bf v}_\bk$ are the neutrino and fermion
velocities, with momentum $\bp$ and $\bk$, respectively. In the last
equality we also assumed isotropy of the background distribution.

As far as the tree-level neutral-current interaction in
Eq.~(\ref{hamgen}) is concerned, $\nu_\mu$, $\nu_\tau$, and $\nu_e$
receive a common potential shift described by the flavour-independent
result of Eq.~(\ref{shift}), which is irrelevant for neutrino
oscillations.  Note also that, in a neutral medium, the sum over all
species cancels the terms proportional to the electric charge $Q_f$ in
Eq.~(\ref{VAcouplings}).

\subsection{Radiative correction for a background of charged fermions}
\label{botellacheck}

The effect we are interested in is the flavour-dependent part of the
loop contribution to the neutral-current couplings $c_{V/A}^{f}$ in
Eq.~(\ref{hamgen}). Technically, the matching of the effective
Hamiltonian coefficients will be obtained by computing the
$\nu_l$--$f$ forward scattering amplitude.  Note also that the
momenta of the scattering particles will be set to zero, since their
typical values in SNe are very small compared to the relevant mass
scales $m_\tau$ and $M_W$. In this approximation all one-loop
integrals reduce to tadpoles.

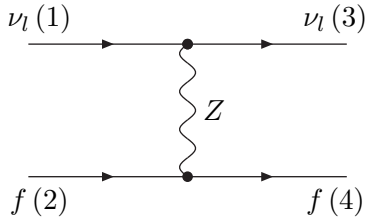
\begin{figure}[!t]
\begin{center}
\begin{picture}(140,80)
\put(0,0){\diagTree}
\end{picture}
\end{center}
\caption{
At tree level, neutrino forward scattering in
a charged-fermion background ($f=u,d,e$)
proceeds through $t$-channel exchange of $Z$ bosons.
\label{treediag}}
\end{figure}

At tree level, $\nu_l$ forward scattering in a charged-fermion
background is described by the Feynman diagram in
Fig.~\ref{treediag}.
The corresponding amplitude is
\begin{eqnarray}
{\mathcal A}^{\nu_l f}_1& =&\frac{- i\,g^2}{4\,M_W^2}
[\ubar_3\,\gamma^\mu \wl \, u_1]
[\ubar_4 \gamma_\mu (c_V^f-c_A^f\gamma_5) u_2]
= \frac{- i\,g^2}{4\,M_W^2}{\mathcal W}_{31}{\mathcal C}^f_{42}\,,
\label{a0}
\end{eqnarray}
where
$\omega_L = (1-\gamma_5)/2$ is the left-handed projector
and we have introduced the currents
\begin{eqnarray}
{\mathcal C}^f_{ij}&=&
\ubar_i \gamma_\mu (c_V^f-c_A^f\gamma_5) u_j\,,
\qquad
{\mathcal W}_{ij}=
{\mathcal C}^{\nu_l}_{ij}=
\ubar_i\,\gamma^\mu \wl \, u_j
\,\label{eq:currents}\,,
\end{eqnarray}
where the Lorentz indices and their contraction are implicitly
understood. At one loop, neutrino-flavour dependent contributions
result only from diagrams where a charged gauge boson ($W^\pm$) or
would-be-Goldstone boson ($\phi^\pm$) couples to a neutrino line,
giving rise to a virtual lepton. In practice we will restrict
ourselves to this type of diagrams and we will retain only terms
that depend on the mass of the lepton and are thus sensitive to its
flavour. Flavour-independent contributions will be systematically
neglected. Note that here and throughout we work in the
't~Hooft-Feynman gauge, in which $M_\phi=M_W$.

As we will see, it is convenient to classify the loop diagrams
according to their topology and to treat box and non-box diagrams
separately. The latter represent the corrections to the
$\nu_l\bar\nu_lZ$ vertex and comprise the vertex diagrams of
Figs.~\ref{vertex1}--\ref{vertex2} as well as the selfenergy
insertions of Fig.~\ref{leg}.
\begin{figure}[!ht]
\begin{center}
\begin{picture}(320,100)
\put(0,0){\diagIa}
\put(180,0){\diagIb}
\end{picture}
\end{center}
\caption{Flavour-dependent vertex corrections
involving $Zl\bar l$ interactions. \label{vertex1}}
\end{figure}
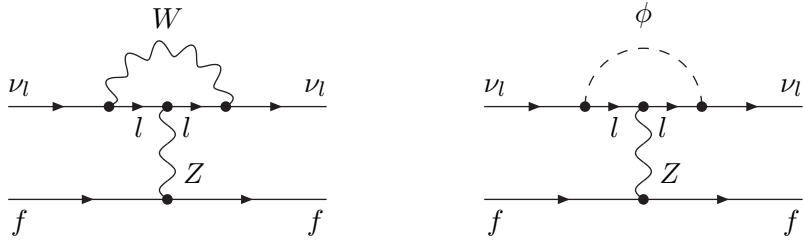%
\begin{figure}[!ht]
\begin{center}
\begin{picture}(420,100)
\put(0,0){\diagIIa}
\put(140,0){\diagIIb}
\put(280,0){\diagIIc}
\end{picture}
\end{center}
\caption{Flavour-dependent vertex corrections
involving $ZW^+W^-$, $Z\phi^+\phi^-$, and $ZW^+\phi^-$ interactions.
In the latter case (rightmost diagram), an additional
contribution with $W\leftrightarrow \phi$ must be taken into account.}\label{vertex2}
\end{figure}
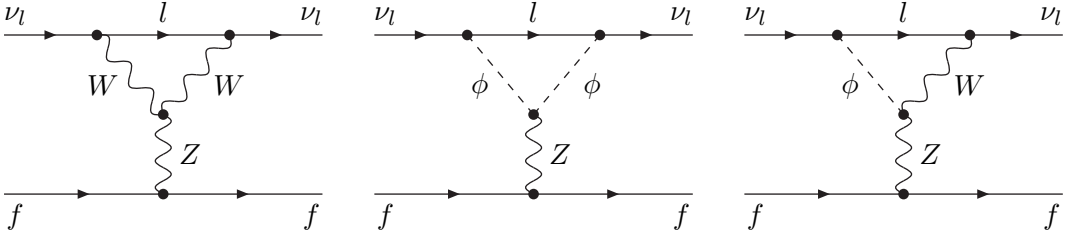%
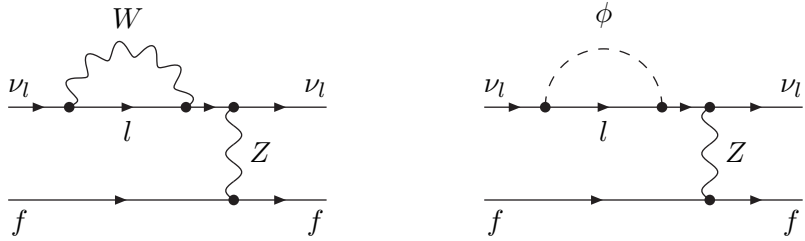
\begin{figure}[!ht]
\begin{center}
\begin{picture}(320,100)
\put(0,0){\diagIIIa}
\put(180,0){\diagIIIb}
\end{picture}
\end{center}
\caption{Flavour-dependent selfenergy insertions in the
incoming neutrino line.
Analogous diagrams for the outgoing neutrino must be considered.
These diagrams have to be understood as on-shell
renormalization factors for the neutrino wave functions.
}\label{leg}
\end{figure}%
Note that all diagrams of non-box type are independent of the nature
of the background fermions $f$ in the sense that the tree-level
$f\bar f Z$ vertex factorizes. Box diagrams can involve two charged
bosons ($WW$, $W\phi$ or $\phi\phi$) and are suppressed unless  both
of them are $W$ bosons. Depending on the isospin nature of the
background  fermions, different box topologies contribute
(Fig.~\ref{box}): in presence of a down-type background ($f=e,d$)
only ladder boxes appear, while crossed boxes contribute only if the
background fermions are of up type ($f=u$ and, for later
application, also $f=\nu$).
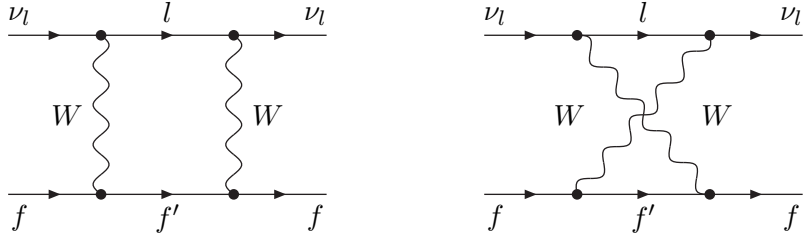
\begin{figure}[!ht]
\begin{center}
\begin{picture}(320,100)
\put(0,0){\diagIVa}
\put(180,0){\diagIVb}
\end{picture}
\end{center}
\caption{Flavour-dependent box corrections.
The ladder box diagrams (left)
contribute only for down-type ($f=e,d$) background fermions,
whereas the crossed box diagrams (right) contribute only for
up-type ($f=u,\nu$) fermions. Box diagrams involving $\phi^\pm$ exchange
are suppressed. }\label{box}
\end{figure}%

Apart from flavour-independent additive terms, the relevant amplitudes for
the diagrams in Fig.~\ref{vertex1} are
\begin{eqnarray}
{\mathcal A}^{\nu_l f}_{2,W} &=& + \frac{i\,g^4}{4\,(4\pi)^2}
\frac{m_l^2}{M_W^4} \bigg[\frac{1}{2}+\sw^2+ \log\frac{m_l^2}{M_W^2}\bigg] \,{\mathcal W}_{31}{\mathcal
C}^f_{42}  \, ,\label{ampli1}\\
{\mathcal A}^{\nu_l f}_{2,\phi}  &=&
-\frac{i\,g^4}{4(4\pi)^2}\frac{m_l^2}{M_W^4}\frac{\sw^2}{2}
\left[\Delta+\frac{1}{2}+\log\frac{\mu^2}{M_W^2}\right]\,{\mathcal W}_{31}{\mathcal
C}^f_{42}\,.
\end{eqnarray}
For the diagrams in Fig.~\ref{vertex2} one finds
\begin{eqnarray}
{\mathcal A}^{\nu_l f}_{3,WWZ} &=&    3\,\cw^2 \frac{i\,g^4}{4 (4 \pi)^2} \frac{m_l^2}{M_W^4}\,{\mathcal W}_{31}{\mathcal
C}^f_{42}\,,\\
{\mathcal A}^{\nu_l f}_{3,\phi\phi Z}  &=&
(\sw^2-\cw^2)\frac{i\,g^4}{4(4\pi)^2}\frac{m_l^2}{M_W^4}
\left( \frac{1}{8}+\frac{\Delta}{4}+ \frac{1}{4}\log\frac{\mu^2}{M_W^2} \right)\,
{\mathcal W}_{31}{\mathcal C}^f_{42}\,,\\
{\mathcal A}^{\nu_l f}_{3,W\phi Z} &=&2\,\sw^2 \frac{i\,g^4 }{4(4\pi)^2 }
\frac{m_l^2}{M_W^4}\,{\mathcal W}_{31}{\mathcal C}^f_{42}\,,
\label{ampli2}
\end{eqnarray}
where in the last equation we
included a factor two corresponding to a second diagram with
$W\leftrightarrow\phi$. Finally, the relevant amplitude for the diagrams in Fig.~\ref{leg} is
\begin{eqnarray}
{\mathcal A}^{\nu_l f}_{4} &=& -  \frac{i\,g^4}{4 (4\pi)^2}
\frac{m_l^2}{M_W^4}
\left(\frac{3}{8}-\frac{\Delta}{4}-\frac{1}{4}
\log\frac{\mu^2}{M_W^2} \right)\,{\mathcal W}_{31}{\mathcal C}^f_{42}\,,
\label{ampli3}
\end{eqnarray}
where, again, we included a factor two for the correction to the
second neutrino. The expression $\Delta = \frac{2}{4-D}-\gamma_E +
\log 4\pi$ represents ultraviolet poles within dimensional
regularization and $\mu$ is the corresponding mass scale. The
complete non-box contribution, i.e.~the sum of the amplitudes
Eqs.~(\ref{ampli1})--(\ref{ampli3}), is free from ultraviolet
divergences and reads
\begin{equation}
{\mathcal A}^{\nu_l f}_{234}=\frac{i\,g^4}{4(4\pi)^2} \frac{m_l^2}{M_W^4} \bigg[3+ \log\frac{m_l^2}{M_W^2}\bigg]\,{\mathcal W}_{31}{\mathcal
C}^f_{42}\, .\label{samenuampl}
\end{equation}For the box diagrams in Fig.~5,
neglecting contributions of order $m_{f'}/M_W$, one obtains
\begin{eqnarray}
{\mathcal A}^{\nu_l d}_{5} &=&
{\mathcal A}^{\nu_l e}_{5} =
\frac{i\,g^4}{(4\pi)^2} \frac{m_l^2}{M_W^4}\left[ \log\left(\frac{m_l^2}{M_W^2}\right) +1\right]
 \,{\mathcal W}_{31}{\mathcal W}_{42}\,,\\
{\mathcal A}^{\nu_l u}_{5} &=&
-\frac{1}{4}\frac{i\,g^4}{(4\pi)^2} \frac{m_l^2}{M_W^4}\left[ \log\left(\frac{m_l^2}{M_W^2}\right) +1\right]
 \,{\mathcal W}_{31}{\mathcal W}_{42}\,.
\label{boxnuu}
\end{eqnarray}
The results for down- and up-type fermions correspond to ladder and
crossed box topologies, respectively.

These
 corrections can be described as flavour-dependent shifts,
\begin{equation}
c_{V/A}^f\to c_{V/A}^{f}+\Delta c_{V/A}^{\nu_lf}\,,
\end{equation}
of fermion couplings in the effective Hamiltonian density ${\mathcal
H}_{\nu_l f}$ of Eq.~(\ref{hamgen}). The shifts induced in the
vector and axial couplings of down- and up-type fermions are
\begin{eqnarray}
\Delta c_{V/A}^{\nu_l d}&=&- \frac{\,g^2}{(4\pi)^2}
\frac{m_l^2}{M_W^2}\bigg[\left(3\,c_{V/A}^{d}+2\right)+\left(c_{V/A}^{d}+2\right)
\log\frac{m_l^2}{M_W^2}\bigg]\,\label{cvshift1} ,\\
\Delta c_{V/A}^{\nu_l u}&=&- \frac{\,g^2}{(4\pi)^2}
\frac{m_l^2}{M_W^2}\bigg[\left(3c_{V/A}^{u}-\frac{1}{2}\right)+\left(c_{V/A}^{u}-\frac{1}{2}\right)
\log\frac{m_l^2}{M_W^2}\bigg]\,.\label{cvshift2}
\end{eqnarray}
Here the terms proportional to $c_{V/A}^f$ originate from non-box
diagrams and the remnant is due to boxes. The relevant shift for the
electron component of the background is obtained by replacing $d\to
e$ in Eq.~(\ref{cvshift1}).

Taking into account that in neutral, ordinary matter $n_e=n_p$,
$n_u=2\,n_p+n_n$, $n_d=n_p+2\,n_n$ and using $n_B= n_p+n_n$, one
obtains the $\nu_\mu$--$\nu_\tau$ energy shift
\begin{equation}
\Delta E_{\nu_\tau\nu_\mu}=\sqrt{2}\,G_{\rm F}\,\sum_f\,\Delta c_V^{\nu_\tau f}\,n_f=3\sqrt{2}\,G_{\rm F}\frac{\,g^2}{(4\pi)^2}
\frac{m_\tau^2}{M_W^2}\left[\log\left(\frac{M_W^2}{m_\tau^2}\right)-1+\frac{Y_n}{3}\right]\,n_B\,,
\end{equation}
equivalent to Eq.~(\ref{eq:botella0}) and in full agreement with
Ref.~\cite{Botella:1986wy}.

\subsection{Extension to neutrinos as background fermions}\label{newresult}
Now we turn to our main case of interest, namely when the
propagating neutrinos ($\nu_l$) interact with a neutrino background
($f=\nu_{l^\prime}$). At tree level the relevant interactions are
described by the four-neutrino neutral-current effective Hamiltonian
\begin{equation}
{\mathcal H}_{\nu\nu}=
\sum_{l,l^\prime}
{\mathcal H}_{\nu_{l}\nu_{l^\prime}}=
\frac{G_{\rm F}}{\sqrt{2}}\sum_{l,l^\prime}
\left[ \bar{\nu}_l\gamma_\alpha \omega_L\nu_l\right]
\left[\bar{\nu}_{l'}\gamma^\alpha \omega_L\nu_{l'}\right]\,
\label{HncNU},
\end{equation}
which is of the same form as Eq.~(\ref{hamgen}) upon identification
of the fermion couplings with $c_V^{\nu_{l'}}=c_A^{\nu_{l'}}=1/2$
and noting that each term with $l\neq l'$ enters twice in the sum. Note
also that Eq.~(\ref{HncNU}) has a $U(3)$ symmetry in flavour space.

The general form of the one-loop effective Hamiltonian is
\begin{equation}
{\mathcal H}_{\nu\nu}=
\sum_{l,l^\prime}
{\mathcal H}_{\nu_l\nu_{l'}}=
\frac{G_{\rm F}}{\sqrt{2}}
\sum_{l,l^\prime}
(1+\kappa^{\nu_l\nu_{l'}})
\left[\bar{\nu}_l\gamma_\alpha \omega_L \nu_l\right]
\left[\bar{\nu}_{l'}\gamma^\alpha \omega_L \nu_{l'}\right]\,
\label{HncNU2}.
\end{equation}
Again only flavour-dependent contributions $\kappa^{\nu_l\nu_{l'}}$
are taken into account. In practice we  consider only terms involving
$m_l^2/M_W^2$ and/or  $m_{l'}^2/M_W^2$. Moreover, the electron and the muon
masses are neglected, i.e.~we use
\begin{equation}
\frac{m_{l}^2}{M_W^2}=\delta_{l\tau}\frac{m_\tau^2}{M_W^2}.
\end{equation}
This implies
\begin{equation}
\kappa^{\nu_e\nu_e}=
\kappa^{\nu_e\nu_\mu}=
\kappa^{\nu_\mu\nu_e}=
\kappa^{\nu_\mu\nu_\mu}=0,
\end{equation}
and one-loop terms proportional to $m^2_\tau/M_W^2$ contribute only
if either $\nu_l$ or $\nu_{l'}$ is a $\nu_\tau$.

We first consider the correction
$\kappa^{\nu_\tau\nu_\beta}=\kappa^{\nu_\beta\nu_\tau}$ with
$\beta\neq\tau$. This case can be easily related to the
charged-fermion background results of Sec.~\ref{botellacheck}. To
this end we can regard $\nu_\tau$ as a neutrino that propagates in a
background with $f=\nu_\beta$. The one-loop diagrams that give rise
to terms of order $m_\tau^2/M_W^2$ are exactly the same as in
Sec.~\ref{botellacheck}:
\begin{itemize}
\item In principle, in addition to the vertices and selfenergies
    of Figs.~\ref{vertex1}--\ref{leg}, which contribute to the
    $\nu_\tau\bar\nu_\tau Z$ vertex, diagrams corresponding to
    the $\nu_\beta\bar\nu_\beta Z$ vertex should be included.
    However, for $\beta\neq \tau$, such diagrams are free from
    $m_\tau^2/M_W^2$ contributions and can thus be neglected.
\item Concerning the box diagrams of Fig.~\ref{box}, the terms
    of order $m_{f'}^2/M_W^2$ that we have neglected in
    Sec.~\ref{botellacheck} correspond to terms of order
    $m_\beta^2/M_W^2$ and remain negligible.
\end{itemize}
This implies that $\kappa^{\nu_\tau \nu_{\beta}}$ can be obtained
from Eq.~(\ref{cvshift2}) for up-type fermions by simply replacing
$u\to\nu_\beta$. By comparing the corresponding Hamiltonians
Eqs.~(\ref{hamgen}) and~(\ref{HncNU2}) we easily see that the
relevant relation~is
\begin{equation}
\kappa^{\nu_\tau\nu_\beta}=\frac{\Delta c_{V/A}^{\nu_\tau\nu_\beta}}{
c_{V/A}^{\nu_\beta}}=2\,\Delta c_{V/A}^{\nu_\tau\nu_\beta}\,.
\end{equation}
Inserting the explicit neutrino couplings we obtain
\begin{eqnarray}
\kappa^{\nu_\tau\nu_\beta}
&=&
\kappa^{\nu_\beta\nu_\tau}
=
-2\,\frac{g^2}{(4\pi)^2}\frac{m_\tau^2}{M_W^2}=-\epsilon\,,
\end{eqnarray}
where $\epsilon$ was defined in Eq.~(\ref{epsi}).

\begin{figure}[!ht]
\begin{center}
\begin{picture}(320,100)
\put(0,0){\diagTreeNNt}
\put(180,0){\diagTreeNNu}
\end{picture}
\end{center}
\caption{At tree level, neutrino forward scattering in
a neutrino background proceeds through $t$-channel exchange of $Z$ bosons (left diagram),
plus $u$-channel exchange in the case of identical flavours only (right diagram).}\label{uchannel}
\end{figure}
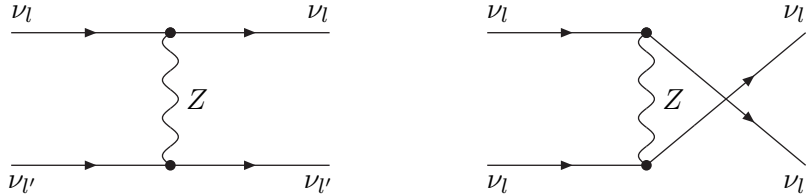%

Finally we consider the equal-flavour corrections
$\kappa^{\nu_\tau\nu_\tau}$ where the following differences must be
taken into account:
\begin{itemize}
\item  All scattering amplitudes  are a factor two larger
    because in addition to the $t$-channel $Z$-exchange diagram
    that contributes to the $\nu_l\neq \nu_{l'}$ interactions
    (left diagram in Fig.~\ref{uchannel}), also a $u$ channel
    opens (right diagram in Fig.~\ref{uchannel}). Moreover, due
    to a Fierz identity, the amplitudes of the $t$- and
    $u$-channel diagrams are identical. In particular, the
    tree-level $\nu_\tau$--$\nu_\tau$ amplitude reads
    [cf.~Eq.~(\ref{a0})]
\begin{eqnarray}
{\mathcal A}^{\nu_\tau \nu_\tau}_1&=&
2\times\left.{\mathcal A}^{\nu_\tau f}_1\right|_{f\to\nu_\tau}=
\frac{- i\,g^2}{2\,M_W^2}{\mathcal W}_{31}{\mathcal
W}_{42}\,.\label{a0TT}
\end{eqnarray}

\item In addition to the vertices and selfenergies
of Figs.~\ref{vertex1}--\ref{leg},
similar diagrams must be included, that describe the correction to the
 $f\bar f Z$ vertex with $f=\nu_\tau$. In practice these two sets of
``non-box" diagrams provide exactly the same correction.
Including a further factor of two due to the presence of $u$-channel
contributions we have [cf.~Eq.~(\ref{samenuampl})]
\begin{equation}
{\mathcal A}^{\nu_\tau\nu_\tau}_{234}
=2\times2\times
\left.{\mathcal A}^{\nu_\tau  f}_{234}\right|_{f\to\nu_\tau}
=\frac{i\,g^4}{(4\pi)^2} \frac{m_l^2}{M_W^4} \bigg[3+ \log\frac{m_l^2}{M_W^2}\bigg]\,{\mathcal W}_{31}{\mathcal
W}_{42}\,.\label{samenuamplNN}
\end{equation}

\item Finally the contribution of the
box diagrams of Fig.~\ref{box} cannot be inferred from the
calculation performed for the matter-background case since the terms
of order  $m_{f'}^2/M_W^2$ cannot be neglected when $f=\nu_{l'}=\nu_\tau$ and $f'=l'=\tau$.
In this case, including again a factor two due to
$u$-channel contributions, the ladder box diagram yields [cf.~Eq.~(\ref{boxnuu})]
\begin{equation}
{\mathcal A}_{5}^{\nu_\tau\nu_\tau}= -\frac{1}{2}\frac{i\,g^4}{(4\pi)^2} \frac{m_\tau^2}{M_W^4}\left[3+2 \log\left(\frac{m_\tau^2}{M_W^2}\right)\right] \,{\mathcal W}_{31}{\mathcal W}_{42}\, .
\label{samenuampl2}
\end{equation}
\end{itemize}
Combining the above results we obtain
\begin{eqnarray}
\kappa^{\nu_\tau\nu_\tau}
&=&
-3\,\frac{g^2}{(4\pi)^2}\frac{m_\tau^2}{M_W^2}=-\frac{3}{2}\,
\epsilon \,,
\end{eqnarray}
where $\epsilon$ was defined in Eq.~(\ref{epsi}).

In the context of neutrino oscillations (see Sec.~\ref{formalism}) it
proves useful to rewrite our results in a more compact notation:
\begin{equation}
{\mathcal H}_{\nu\nu}=\frac{G_{\rm F}}{\sqrt{2}}
\left\{[\bar{\nu}({\sf 1}-\epsilon\T)\gamma^\lambda\omega_L\nu]
[\bar{\nu}({\sf 1}-\epsilon\T)\gamma_\lambda\omega_L\nu]
+\frac{\epsilon}{2}[\bar{\nu}\T\gamma^\lambda\omega_L\nu]
[\bar{\nu}\T\gamma_\lambda\omega_L\nu]\right\}\,,
\label{hamnew}
\end{equation}
where $\nu$ is now a vector in flavor space and in the
weak-interaction basis
\begin{equation}
{\sf T}={\rm diag}(0,0,1)\,,
\label{TSdef}
\end{equation}
i.e. the matrix $\T$ is a projector on the $\tau$
direction in flavour space. Equation~(\ref{hamnew}) holds at leading
order in $\epsilon$. Note that the renormalization of the
$\nu_\tau$--$\nu_\tau$ coupling implied by the first term in
Eq.~(\ref{hamnew}) is $1-2\epsilon$ and has to be summed to the
second term to provide the correct result $1-3\epsilon/2$. In other
words, not only is the original $U(3)$ symmetry of the tree-level
Hamiltonian Eq.~$(\ref{HncNU})$ broken, but a simple description by
renormalizing the coefficients $c^{\nu}_{V/A}$ is not possible.

For comparison with the refractive energy shifts caused by a
background of charged fermions, we recall that at tree level a
neutrino $\nu_l$ in a homogeneous and isotropic bath of $\nu_{l'}$
experiences
\begin{equation}
{V}_{\nu_l \nu_{l'}}=\sqrt{2}G_{\rm F}\,
(1+\delta_{ll'})\,n_{\nu_{l'}}.\label{velljmath}
\end{equation}
Of course, the usual ultra-relativistic approximation was used where
neutrinos are purely left-handed and $n_{\nu_{l'}}$ is once more a net
density of neutrinos minus antineutrinos. If neither $l$ nor $l'$
are the $\tau$ flavour, radiative corrections are negligible at our
level of approximation. For $l=\tau$ and $l'=\beta\not=\tau$ our
results imply
\begin{equation}
{V}_{\nu_\tau \nu_{\beta}}=\sqrt{2}G_{\rm F}\,(1-\epsilon)\,n_{\nu_{\beta}}
\end{equation}
and analogously under the exchange of
$\tau\leftrightarrow\beta$. Finally, for $l=l'=\tau$ we find
\begin{equation}
{V}_{\nu_\tau \nu_{\tau}}=\sqrt{2}G_{\rm F}\,(2-3\epsilon)\,n_{\nu_{\tau}}\,.
\end{equation}
However, in the context of neutrino oscillations these energy shifts
do not provide a useful understanding of the role of
neutrino--neutrino refraction because the background neutrinos
themselves are, in general, coherent superpositions of weak
interaction states.

\section{Including radiative corrections in the oscillation equations}
\label{formalism}
To see how neutrino--neutrino refraction influences flavour
oscillations it is most economical to represent the ensemble by
flavour-space matrices of occupation numbers $\varrho_{ij}({\bf
  p})=\langle a_i^\dag({\bf p})a_j({\bf p})\rangle$ where $i$ and $j$
are flavour indices, $a^\dag$ and $a$ are creation and annihilation
operators and $\langle{\cdots}\rangle$ is an expectation value for the
ensemble~\cite{Sigl:1992fn, Dolgov:1980cq}. The diagonal elements of
these flavour-space matrices are the usual occupation numbers, whereas
the off-diagonal elements encode relative phases between the different
flavour components. The simultaneous treatment of neutrinos and
antineutrinos becomes particularly transparent if one defines
$\bar\varrho_{ij}({\bf p})=\langle \bar a_j^\dag({\bf p})\bar a_i({\bf
  p})\rangle$ with ``crossed'' flavour indices where all over-barred
quantities refer to antiparticles~\cite{Sigl:1992fn}. The equations of
motion (EOM) for the flavour evolution are
\begin{equation}
\I\dot\varrho_{\bf p}
=\left[{\sf \Omega}_{\bf p},\varrho_{\bf p}\right]
\quad{\rm and}\quad
\I\dot{\bar\varrho}_{\bf p}
=\left[\bar{\sf \Omega}_{\bf p},\bar\varrho_{\bf p}\right]\,,
\end{equation}
where $[\cdot,\cdot]$ is a commutator and the ``matrices of
oscillation frequencies'' ${\sf \Omega}_{\bf p}$ and $\bar{\sf
\Omega}_{\bf p}$ play the role of Hamiltonian operators in flavour
space for the evolution of mode ${\bf p}$.

In an isotropic ensemble, vacuum masses, matter, and neutrinos
provide at tree level the usual terms~\cite{Sigl:1992fn}
\begin{equation}\label{eq:hamilt}
{\sf \Omega}_{\bf p}=
+\frac{{\sf M}^2}{2E}
+\sqrt{2}G_{\rm F}\left({\sf N}_l+{\sf N}_\nu\right)\,,\:\:\:\:\:\:\:\:\bar{\sf \Omega}_{\bf p}=
-\frac{{\sf M}^2}{2E}
+\sqrt{2}G_{\rm F}\left({\sf N}_l+{\sf N}_\nu\right)\,,
\end{equation}
where ${\sf M}^2$ is the neutrino mass-squared matrix and $E=|{\bf
  p}|$ for ultrarelativistic neutrinos. In the weak interaction basis,
${\sf M}^2=U\,D\,U^\dagger$ where $D={\rm diag}(m_1^2,m_2^2,m_3^2)$
and $U$ is the usual mixing matrix.  The matrix of number densities
${\sf N}_l$ for charged leptons is defined in the weak-interaction
basis with diagonal elements \hbox{$n_l= n_{l^-}-n_{l^+}$}.  The
corresponding neutrino matrix is
\begin{equation}
{\sf N}_\nu=\int\frac{\D^3{\bf q}}{(2\pi)^3}\,
\left(\varrho_{\bf q}-\bar\varrho_{\bf q}\right)\,.
\end{equation}
If the medium is not isotropic, as in the case for neutrinos streaming
from a SN core, the usual factor $(1-{\bf v}_{\bf q}\cdot{\bf v}_{\bf
  p})$ must be included under the integral and the matrix ${\sf
  N}_\nu$ then also depends on ${\bf p}$. This structure is identical
in all that follows, so without loss of generality we can study the
radiative modification assuming an isotropic medium.

To see how our radiative corrections modify the structure of ${\sf
  N}_\nu$ it is easiest to begin with a hypothetical case of
non-universal neutrino interactions of the form
\begin{equation}
{\mathcal H}_{\sG}=\frac{G_{\rm F}}{\sqrt{2}}
(\bar{\nu}\sG\gamma^\lambda\omega_L\nu)
(\bar{\nu}\sG\gamma_\lambda\omega_L\nu)\,,
\label{hamSigl}
\end{equation}
where $\sG$ is a matrix in flavour space of dimensionless coupling
constants. The tree-level standard-model case is
$\sG={\sf 1}$ (unit matrix in flavour space). It was previously shown
that the oscillation equation is to be modified in the
form~\cite{Sigl:1992fn}
\begin{equation}
{\sf N}_\nu\to
{\sf N}_\nu^{\rm eff}(\sG)
=\int\frac{\D^3{\bf q}}{(2\pi)^3}\,
\big\{
\sG\left(\varrho_{\bf q}-\bar\varrho_{\bf q}\right)\sG+\sG\,\textrm{Tr}[
(\varrho_{\bf q} -{\bar \varrho}_{\bf q})\sG]\big\}\,.
\label{nunurho2}
\end{equation}
For $\sG={\sf 1}$ this expression is identical to ${\sf N}_\nu$ up to
an additive term that is proportional to the unit matrix and to the
net total neutrino density
\begin{equation}
n_\nu=\int\frac{\D^3{\bf q}}{(2\pi)^3}\,
\textrm{Tr}(\varrho_{\bf q}-\bar\varrho_{\bf q})\,.
\label{eq:trace}
\end{equation}
This quantity is conserved under oscillations, so ${\sf N}_\nu$ and
${\sf N}_\nu^{\rm eff}({\sf 1})$ play identical roles and actually we
could subtract Eq.~(\ref{eq:trace}) from Eq.~(\ref{nunurho2}) to
achieve ${\sf N}_\nu={\sf N}_\nu^{\rm eff}({\sf 1})$.

Our calculation of the radiative correction of neutrino--neutrino
interactions reveals that the flavour-sensitive difference of the
scattering amplitudes can be expressed, at leading order and in the
``large $m_\tau$ limit,'' in the form of Eq.~(\ref{hamnew}).
Therefore, radiative corrections effectively enter with two
different matrices of non-universal couplings. In the oscillation
equation we thus need to substitute
\begin{equation}
{\sf N}_\nu\to {\sf N}_\nu^{\rm eff}({\sf 1}-\epsilon\T)
+\frac{\epsilon}{2}\, {\sf N}_\nu^{\rm eff}(\T)\,,
\label{resEOM}
\end{equation}
where the projection matrix $\T$ was defined in Eq.~(\ref{TSdef}).
In the weak-interaction basis this is explicitly to leading order in
$\epsilon$
\begin{equation}
{\sf N}_\nu\to {\sf N}_\nu
-\epsilon\left(
\begin{array}{ccc}
0 & 0 & {\sf N}_\nu^{e\tau}\\
0 & 0 & {\sf N}_\nu^{\mu\tau}\\
{\sf N}_\nu^{\tau e} & ~~~{\sf N}_\nu^{\tau\mu}~~~ &
{\sf N}_\nu^{ee}+{\sf N}_\nu^{\mu\mu}+2{\sf N}_\nu^{\tau\tau}
\end{array}
\right)\,,
\end{equation}
where we have neglected a term proportional to the unit matrix that
is irrelevant for oscillations.

\section{Conclusions}  \label{concl}

We have calculated the flavour non-universal correction for
neutrino--neutrino refraction in the framework of the Standard Model.
The dominant term arises at loop level and involves $\tau$ leptons
circulating in the loop. In the course of our derivation we have
reproduced a similar term for a background medium of charged
fermions that had been calculated previously by Botella, Lim and
Marciano~\cite{Botella:1986wy}. One novel feature of our result is
that the radiatively corrected effective neutrino--neutrino
Hamiltonian can not be expressed in terms of renormalized tree-level
coupling constants. This different structure derives from the box
diagrams with the exchange of two W bosons.

The impact of neutrino--neutrino refraction on collective neutrino
oscillations is not easily assessed by comparing refractive energy
shifts relative to those caused by an ordinary-matter background.
The recent torrent of activities concerning collective SN neutrino
oscillations was essentially triggered by the insight that the
nonlinear nature of the equations allows for large collective
effects even if the ordinary-matter background causes larger energy
shifts than the neutrino background. This is traced back to the
phenomenon that the refractive-index matrix caused by ordinary
matter and by background neutrinos are usually not diagonal in the
same basis because of neutrino flavor oscillations.

Our radiative corrections provide non-universal neutrino--neutrino
interactions similar to, but not identical with, the case of
non-standard interactions studied in Ref.~\cite{Blennow:2008er},
suggesting the possibility of interesting collective oscillation
phenomena. However, a realistic assessment requires a dedicated
study including the impact of a dense ordinary-matter background and
multi-angle effects.

We also note that in certain supersymmetric scenarios the
neutrino--fermion radiative corrections could be enhanced up to two
order of magnitudes with respect to the standard
case~\cite{Roulet:1995qb,wsusy}. Similar enhancements might be expected
for neutrino--neutrino refraction.


\begin{acknowledgments}
In Munich, we acknowledge partial support by the Deutsche
Forschungsgemeinschaft under grant TR-27 ``Neutrinos and Beyond''
and the Cluster of Excellence ``Origin and Structure of the
Universe.'' We thank B. Dasgupta and J. Gava for comments. The work of AM is
supported by the Italian Istituto Nazionale di Fisica Nucleare
(INFN). P.S. thanks the Galileo Galilei Institute for Theoretical
Physics for hospitality and the INFN for partial support during the
initial stage of this work. A.M. thanks CERN for kind hospitality
during the initial development of this project.
\end{acknowledgments}



\begin{thebibliography}{00}

\bibitem{Wolfenstein:1977ue}
  L.~Wolfenstein,
  ``Neutrino oscillations in matter,''
  Phys.\ Rev.\  D {\bf 17}, 2369 (1978).

\bibitem{Mikheev:1986gs}
  S.~P.~Mikheev and A.~Yu.~Smirnov,
  ``Resonance enhancement of oscillations in matter and solar neutrino
  spectroscopy,''
  Yad.\ Fiz.\  {\bf 42}, 1441 (1985)
  [Sov.\ J.\ Nucl.\ Phys.\  {\bf 42}, 913 (1985)].

\bibitem{Botella:1986wy}
  F.~J.~Botella, C.~S.~Lim and W.~J.~Marciano,
 ``Radiative corrections to neutrino indices of refraction,''
  Phys.\ Rev.\  D {\bf 35}, 896 (1987).

\bibitem{Dighe:1999bi}
  A.~S.~Dighe and A.~Y.~Smirnov,
  ``Identifying the neutrino mass spectrum from the neutrino burst
  from a supernova,''
  Phys.\ Rev.\  D {\bf 62}, 033007 (2000)
  [hep-ph/9907423].


\bibitem{Akhmedov:2002zj}
  E.~K.~Akhmedov, C.~Lunardini and A.~Y.~Smirnov,
  ``Supernova neutrinos:
  Difference of $\nu_\mu$--$\nu_\tau$ fluxes and conversion
  effects,''
  Nucl.\ Phys.\  B {\bf 643}, 339 (2002)
  [hep-ph/0204091].

\bibitem{Pantaleone:1992eq}
  J.~T.~Pantaleone,
  ``Neutrino oscillations at high densities,''
  Phys.\ Lett.\  B {\bf 287}, 128 (1992).

\bibitem{Sigl:1992fn}
  G.~Sigl and G.~Raffelt,
  ``General kinetic description of relativistic mixed neutrinos,''
  Nucl.\ Phys.\  B {\bf 406}, 423 (1993).

\bibitem{Kostelecky:1994dt}
  V.~A.~Kostelecky and S.~Samuel,
  ``Selfmaintained coherent oscillations in dense neutrino gases,''
  Phys.\ Rev.\  D {\bf 52}, 621 (1995).
  [hep-ph/9506262].

\bibitem{Pastor:2001iu}
  S.~Pastor, G.~G.~Raffelt and D.~V.~Semikoz,
  ``Physics of synchronized neutrino oscillations caused by
  self-interactions,''
  Phys.\ Rev.\  D {\bf 65}, 053011 (2002).
  [hep-ph/0109035].

\bibitem{Duan:2009cd}
  H.~Duan and J.~P.~Kneller,
  ``Neutrino flavor transformation in supernovae,''
  arXiv:0904.0974 [astro-ph.HE].

\bibitem{EstebanPretel:2007yq}
  A.~Esteban-Pretel, S.~Pastor, R.~Tom{\`a}s, G.~G.~Raffelt and G.~Sigl,
  ``Mu-tau neutrino refraction and collective three-flavor
  transformations in supernovae,''
  Phys.\ Rev.\  D {\bf 77}, 065024 (2008)
  [arXiv:0712.1137].

\bibitem{Gava:2008rp}
 J.~Gava and C.~Volpe,
 ``Collective neutrinos oscillation in matter and CP-violation,''
 Phys.\ Rev.\  D {\bf 78}, 083007 (2008)
 [arXiv:0807.3418].
 
\bibitem{EstebanPretel:2008ni}
  A.~Esteban-Pretel, A.~Mirizzi, S.~Pastor,
  R.~Tom{\`a}s, G.~G.~Raffelt, P.~D.~Serpico and G.~Sigl,
  ``Role of dense matter in collective supernova neutrino
  transformations,''
  Phys.\ Rev.\  D {\bf 78}, 085012 (2008)
  [arXiv:0807.0659].

\bibitem{Raffelt:1996wa}
  G.~G.~Raffelt,
``Stars As Laboratories For Fundamental Physics: The Astrophysics Of
Neutrinos, Axions, And Other Weakly Interacting Particles,''
 Chicago, USA: Univ. Pr. (1996) 664 p.

\bibitem{Fukugita:2003en}
  M.~Fukugita and T.~Yanagida,
 ``Physics of neutrinos and applications to astrophysics,''
Berlin, Germany - Springer (2003) 593 p.

\bibitem{Giunti:1990pp}
  C.~Giunti, C.~W.~Kim and W.~P.~Lam,
  ``Radiative decay and magnetic moment of neutrinos in matter,''
  Phys.\ Rev.\  D {\bf 43}, 164 (1991).

\bibitem{Dolgov:1980cq}
  A.~D.~Dolgov,
  ``Neutrinos in the early universe,''
  Yad.\ Fiz.\  {\bf 33} (1981) 1309
  [Sov.\ J.\ Nucl.\ Phys.\  {\bf 33}, 700 (1981)].

\bibitem{Blennow:2008er}
  M.~Blennow, A.~Mirizzi and P.~D.~Serpico,
  ``Nonstandard neutrino--neutrino refractive effects in dense
  neutrino gases,''
  Phys.\ Rev.\  D {\bf 78}, 113004 (2008)
 [arXiv: 0810.2297].

\bibitem{Roulet:1995qb}
  E.~Roulet,
  ``Supersymmetric radiative corrections to neutrino indices of refraction,''
 Phys.\ Lett.\  B {\bf 356}, 264 (1995)
  [hep-ph/9506221].
  
\bibitem{wsusy}
  J.~Gava and C.~C.~Jean-Louis,
``SUSY radiative corrections on mu-tau neutrino refraction including possible
R-parity breaking interactions,''
  arXiv:0907.3947 [hep-ph].

\end{thebibliography}
\end{document}